\documentclass[12pt,epsfig,a4]{article} 
\newcommand{\vs}{\vspace{-0.25cm}}
\usepackage{amsmath,graphicx,wasysym}
\begin{document} 
\begin{center}
{\Large{\bf Loop functions of sunset diagrams in 2+1 space-time dimensions}}  

\bigskip

 N. Kaiser \\
\medskip
{\small Physik-Department T39, Technische Universit\"{a}t M\"{u}nchen,
   D-85747 Garching, Germany\\

\smallskip

{\it email: nkaiser@ph.tum.de}}
\end{center}
\medskip
\begin{abstract}
In these notes the relativistic $n$-body phase-phase is calculated iteratively in $2+1$ space-time dimensions for all $n$. The obtained result shows a simple power-law behavior $\alpha_n (\mu-M)^{n-2}/\mu$ with a dependence only on the total mass $M=m_1+\dots + m_n$. As a consequence of this feature, the $(n-1)$-loop integrals $J_n(-q^2)$ associated to sunset diagrams with $n$ internal lines can be expressed through elementary (arctangent and logarithmic) functions, modulo polynomial terms in $q^2$ with regularization-dependent coefficients. An outlook to the analogous situation in $4+1$ space-time dimensions is given by computing the $n$-body phase-phases for $n=2,3,4,5$ with their totally symmetric dependence on the involved masses. Moreover, a digression to $1+1$ space-time dimensions reveals that there the three-body phase-space is already proportional to a complete elliptic integral.      
\end{abstract}

\section{Introduction and summary}
The understanding of the structure behind the master-integrals that occur in the evaluation of multi-loop Feynman diagrams is a subject of high relevance in current research in theoretical high-energy physics. Numerous advanced mathematical topics, such as Picard-Fuchs differential equations, elliptic curves, Calabi-Yau manifolds, Hopf algebras, cluster algebras, etc. are encountered in these sophisticated studies \cite{weinzierl}. Moreover, it was found that there exist relations between the sets of master integrals when the space-time dimension $d$ is increased or lowered by two units, $d\to d\pm 2$.

In the present notes the loop functions associated to the sunrise diagrams with $n$ internal lines are analyzed in $2+1$ space-time dimensions. The pertinent imaginary part is the integrated $n$-body phase-space, which exhibits a remarkably simple form,   $\Gamma_n(\mu; m_1,\dots, m_n)=\alpha_n (\mu-M)^{n-2}/\mu$, with a power-law dependence on the energy $\mu$ and a parametric dependence on the masses only through the total sum $M=m_1+\dots + m_n$. As a consequence of this feature the loop-integrals $J_n(-q^2)$ (in the euclidean domain) can    be expressed in terms of elementary (arctangent and logarithmic) functions, modulo polynomial terms in $q^2$ with regularization-dependent coefficients. An outlook to the analogous situation in $4+1$ space-time dimensions is also given by computing the $n$-body phase-phases for $n=2,3,4,5$ with their dependence on elementary  symmetric polynomials of the involved  masses $m_1, \dots, m_n$. One finds $\widetilde\Gamma_n(\mu; m_1,\dots, m_n) =(8\pi)^{3-2n} \mu^{-3}(\mu -M)^{2n-3}P_{n+1}(\mu)$, where the last factor is a polynomial in $\mu$ of degree $n+1$.  Moreover, a digression to $1+1$ space-time dimensions reveals that there the three-body phase-space is already proportional to a complete elliptic integral. This signals a similar complexity of multi-loop integrals as in the physical $3+1$ space-time dimensions.

\begin{figure}[h]\centering\includegraphics[width=0.5\textwidth]{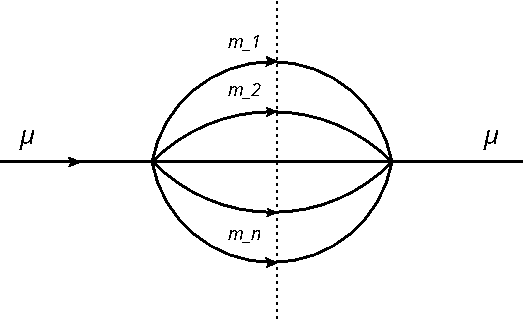}\caption{Sunset diagram with $n$ internal lines carrying masses $m_1, m_2, \dots , m_n$. The vertical dotted line symbolizes the cut, which gives rise to the integrated $n$-body phase-space. }\end{figure}

\section{Integrated two-, three-, and $n$-body phase space}
With the usual normalization factors, the two-body phase-space with final-state particles of masses $m_1$ and $m_2$ is calculated in $2+1$ space-time dimensions as follows: 
\begin{eqnarray}
 \Gamma_2(\mu; m_1, m_2)&=& \int{d^2k_1 \over (2\pi)^2 2\omega_1}\int{d^2k_2 \over (2\pi)^2 2\omega_2} \,(2\pi)^3 \delta^{(3)}\big(\vec P -\vec k_1-\vec k_2\big) \nonumber \\ &=& {1\over 4}\int_0^\infty dk{k \over \sqrt{m_1^2+k^2} \sqrt{m_2^2+k^2}}\, \delta\Big(\mu - \sqrt{m_1^2+k^2}- \sqrt{m_2^2+k^2}\,\Big)\,,\end{eqnarray}
with $\vec P = (\mu, 0,0)$ in the center-of-mass frame. The derivative of the argument of the $\delta$-function with respect to $k$, inserted with its reciprocal value, cancels the other factors of the integrand, and only the center-of-mass energy $\mu = \sqrt{m_1^2+k^2} +\sqrt{m_2^2+k^2}$ remains as a  denominator. Thus one obtains the following simple result for the integrated two-body phase-space in $2+1$ dimensions:
\begin{equation}
\Gamma_2(\mu; m_1, m_2)= {1\over 4 \mu}\,\theta(\mu-m_1-m_2)\,,
\end{equation}
which exhibits a non-zero starting value at threshold $\mu_\text{th} = m_1+m_2$.

The calculation of the three-body phase-space with final state particles of masses $m_1, m_2, m_3$ proceeds as follows:
\begin{eqnarray}
 \Gamma_3(\mu; m_1, m_2,m_3)&=& \int{d^2k_1 \over (2\pi)^2 2\omega_1}\int{d^2k_2 \over (2\pi)^2 2\omega_2} \int{d^2k_3 \over (2\pi)^2 2\omega_3} \,(2\pi)^3 \delta^{(3)}\big(\vec P -\vec k_1-\vec k_2-\vec k_3\big) \nonumber \\ &=& {1\over 16\pi^2}\iint_{D(\omega_1,\omega_2)> 0}\!\!\! d\omega_1 d\omega_2 \big[D(\omega_1,\omega_2)\big]^{-1/2} 
 \,,\end{eqnarray}
where the angular integration has been performed, and the cubic polynomial  $D(\omega_1,\omega_2) = (\mu^2-2 \mu\, \omega_1 +m_1^2)(\omega_2^+ -\omega_2)(\omega_2-\omega_2^-)> 0$ demarcates the kinematically allowed Dalitz region in the $\omega_1\omega_2$-plane. Using $\int_{\omega_2^-}^{\omega_2^+} d\omega_2[(\omega_2^+ -\omega_2) (\omega_2-\omega_2^-) ]^{-1/2} = \pi$ and the upper boundary $\omega_1^\text{end} = (\mu^2+m_1^2-(m_2+m_3)^2)/2\mu$, one arrives at the elementary integral
\begin{equation}\Gamma_3(\mu; m_1, m_2,m_3)= {1\over 16 \pi} \int_{m_1}^{\omega_1^\text{end}}\!\!\! d\omega_1 (\mu^2 -2\mu\, \omega_1 +m_1^2)^{-1/2}\,,
\end{equation}
which has  the solution \begin{equation}\Gamma_3(\mu; m_1, m_2,m_3)={1\over 16\pi \mu}(\mu - m_1-m_2-m_3)  \,,\end{equation}
and the obvious condition $\mu > m_1+m_2+m_3$ must hold. One observes that in $2+1$ dimensions the three-body phase-space grows linearly from threshold, and interestingly it depends only on the total sum $M=m_1+m_2+m_3$ of the three masses.

It is well-known that the integrated phase-space for $n$ final state particles (with masses $m_1,\dots, m_n)$ can be recursively built up from the expression for the two-body phase-space. The formula connecting the integrated phase-spaces for $n$ and $n-1$ particles reads (after adaption of $2\pi$ factors) from ref.\cite{kinemat}:
 \begin{equation}\Gamma_n(\mu; m_1,\dots,m_n)={1\over \pi}
 \int_{m_1+\dots +m_{n-1}}^{\mu - m_n} \!\!\!d\mu' \mu' \, \Gamma_2(\mu; \mu',m_n)\, \Gamma_{n-1}(\mu'; m_1,\dots,m_{n-1})  \,,\end{equation}
with $\mu'$ the total center-of-mass energy of the $(n-1)$-particle subsystem.  Using the result in eq.(2) for the two-body phase-space, one reproduces quickly the just derived expression for the integrated three-body phase-space 
   
\begin{equation}\Gamma_3(\mu; m_1, m_2,m_3)= {1\over 16 \pi} \int_{m_1+m_2}^{\mu - m_3}\!\!d\mu' {\mu' \over \mu\, \mu'} = {1\over 16\pi \mu}(\mu - m_1-m_2-m_3) \,.
\end{equation}
The integrated phase-space for four, five, and more particles in $2+1$ dimensions can be recursively calculated with the help of eq.(6). For $n$ particles the result is given by the power-law expression
 \begin{equation} \Gamma_n(\mu; M) = \alpha_n {(\mu -M)^{n-2} \over \mu}\,,
\end{equation}
with $M=\sum_{j=1}^n m_j$ the total sum of all masses, and the coefficient $\alpha_n$ is determined by a recursion relation as:

\begin{equation} \alpha_n= {\alpha_{n-1} \over 4\pi} {1\over n-2}\,, \qquad \alpha_2 ={1\over 4}\,, \qquad \alpha_n = \big[4(4\pi)^{n-2} (n-2)!\big]^{-1}\,.
\end{equation}
In summary, this leads to an extraordinarily simple result for the integrated $n$-body phase-space in $2+1$ dimensions:
\begin{equation} \Gamma_n(\mu; M) = {(\mu -M)^{n-2} \over 4\mu(4\pi)^{n-2} (n-2)!}\,, \qquad M=\sum_{j=1}^n m_j\,.
\end{equation}

\section{Loop functions of sunset diagrams}
The sunset diagram of order $n$, shown in Fig.\,1,  is a $(n-1)$-loop diagram, where the incoming leg (carrying energy-momentum $\vec P$) splits at one point into $n$ internal lines, corresponding to the propagators of scalar particles with masses $m_1,\dots, m_n$, and then these $n$ lines recombine at one point to the outgoing leg (carrying again energy-momentum $\vec P$). The associated loop function $J_n(s)$ with $s=\vec P\!\cdot\!\vec P$ has an imaginary part for $s>(m_1+\dots + m_n)^2=M^2$ 
that is equal to $1/2$ times the integrated $n$-body phase-space: 
\begin{equation}
\text{Im}J_n(s) = {1\over 2} \Gamma_n(\sqrt{s};M)\,.    \end{equation}  
In the euclidean region $s=-q^2<0$ the loop function is conveniently calculated via the dispersion relation
\begin{equation}
J_n(-q^2) = {1\over \pi}\int_M^\infty d\mu \,{\mu \, \Gamma_n(\mu;M)\over \mu^2 +q^2}\,,    \end{equation} 
but subtractions (at $q^2=0$) are necessary for $n\geq 3$ due to the polynomial growth of $\Gamma_n(\mu;M)$ with $\mu$. The introduces coefficients of a polynomial subtraction term, that need to be calculated in some regularization scheme, like cutoff or dimensional regularization.  

One finds for $n=2$ the one-loop function
\begin{equation}
J_2(-q^2) = {1\over 4\pi q} \arctan{q \over m_1+m_2}\,.    \end{equation} 
This result is also readily obtained with the help of Feynman parametrization as
\begin{equation}
J_2(-q^2) = {1\over 8\pi}\int_0^1 dx \big[m_1^2 x+m_2^2(1-x)+q^2 x(1-x)\big]^{-1/2} = {1\over 4\pi\, q} \arctan{q \over m_1+m_2}\,.    \end{equation} 
With little effort one finds for $n=3$ the two-loop function:
\begin{equation}
J_3(-q^2) -J_3(0)= {1\over 16\pi^2}\bigg\{ 1 - {M\over q }\arctan{q \over M}-{1\over 2} \ln\Big( 1+ {q^2 \over M^2}\Big)\bigg\}\,,    \end{equation} 
with $M=m_1+m_2+m_3$. The divergent subtraction constant $J_3(0)$ is expected to involve the logarithm of $M$, and the computation in $d$  dimensions with a regularization scale $\kappa$ gives:
 \begin{eqnarray}
J_3(0) &=& (4\pi)^{-d} \Gamma(3-d) \int_0^1 dx\int_0^1 dy {y^{1-d/2} H^{d-3} \kappa^{6-2d}\over \big[1+(x-1-x^2)y\big]^{d/2}} \nonumber \\  &=& {1\over 64 \pi^3}\int_0^1 dx\int_0^1 dy {1 \over \sqrt{y} [1+(x-1-x^2)y]^{3/2}} \bigg\{ {1\over 3-d} -\gamma_E +\ln 4\pi -2 \ln {M \over \kappa} \nonumber \\ && \qquad \qquad \qquad\qquad - \ln {H \over M^2} +{1\over 2} \ln\big[y+(x-1-x^2)y^2\big]\bigg\}\,,
\end{eqnarray}
with $H = m_1^2(1-y)+m_2^2(1-x)y +m_3^2 xy$, and one has expanded around $d=3$. The integral $\int_0^1 dy$ for the (four) terms is the second line is performed analytically and it leads to $\int_0^1 dx\, 2(x-x^2)^{-1/2} = 2\pi$. The same contribution, $ \int_0^1 dx\, 2(x-x^2)^{-1/2} =2\pi$, comes as a part after performing analytically the integral   $\int_0^1 dy$ for the (two) terms in the third line, while a remainder with a complicated dependence on the masses $m_1,m_2,m_3$ gives a numerical value very close to zero, at the level of $10^{-14}$. When taking it for granted  that this remainder depends only on the sum $M=m_1+m_2+m_3$, one can set $m_1=m_2=0$ and obtains directly $\int_0^1dx [x(1-x)]^{-1/2} [\ln(1-x)-\ln x]=0$. Thus the subtraction constant $J_3(0)$, split off in eq.(15), has in dimensional regularization the value
\begin{equation}
J_3(0) = {1\over 32 \pi^2} \bigg\{ {1\over 3-d} -\gamma_E +\ln 4\pi -2 \ln {M \over \kappa} +1 \bigg\}\,.
\end{equation}    
 One finds for $n=4$ the three-loop function:   
   \begin{equation}
J_4(-q^2) -J_4(0)= {1\over 2(4\pi)^3}\bigg\{  {M^2-q^2\over q }\arctan{q \over M}-M+M \ln\Big( 1+ {q^2 \over M^2}\Big)\bigg\}\,,    \end{equation} 
with $M=m_1+m_2+m_3+m_4$, and  $J_4(0)$ a yet undetermined subtraction constant that will include the logarithmic piece $(4\pi)^{-3} M \ln(M/\kappa)$.

One finds for $n=5$ the four-loop function: 
\begin{eqnarray}
J_5(-q^2) -J_5(0)+q^2 J_5'(0) &=& {1\over 6(4\pi)^4}\bigg\{  {M\over q }(3q^2-M^2)\arctan{q \over M}\nonumber \\ && +{1\over 2}(q^2-3 M^2) \ln\Big( 1+ {q^2 \over M^2}\Big)+M^2-{11\over 6} q^2\bigg\}\,,   \end{eqnarray}  
with $M=m_1+m_2+m_3+m_4+m_5$. Moreover, $J_5(0)=-{1\over 2}(4\pi)^{-4}M^2 \ln(M/\kappa)+\dots$ and   $J_5'(0)=-{1\over 6}(4\pi)^{-4} \ln(M/\kappa)+\dots$ are two regularization-dependent subtraction constants.

Modulo additive polynomial terms in $q^2$, the result for the $(n-1)$-loop function associated to the sunset diagram (Fig.\,1) reads in $2+1$ dimensions:

\begin{equation}J_n(-q^2) = {1 \over (4\pi)^{n-1} (n-2)!} \bigg\{ A_n(q^2, M) \, {1\over q} \arctan{q \over M}+L_n(q^2,M) \bigg[ \ln\Big( 1+ {q^2 \over M^2}\Big)+2\ln{M\over \kappa}\bigg]+ \text{ polyn}\bigg\} \,,    \end{equation} 
where the polynomials in $q^2$ multiplying the arctangent and logarithmic  function are 
\begin{equation}
A_n(q^2, M) = \text{Re} \big[(i q - M)^{n-2}\big]\,, \qquad L_n(q^2, M) = -{1\over 2q} \text{Im} \big[(i q - M)^{n-2}\big]\,.
\end{equation} 
These expressions are derived from the partial fraction decomposition
\begin{equation}{(\mu -M)^{n-2} \over \mu^2+q^2} = {(i q - M)^{n-2}\over 2iq(\mu - i q)} + {(-i q - M)^{n-2}\over -2iq(\mu + i q)} +\text{polyn}\,,   \end{equation} 
followed by a recombination into terms proportional to $1/(\mu^2+q^2)$ and $-2\mu/(\mu^2+q^2)$, and observing that these two pieces integrate to the basic functions $q^{-1} \arctan(q/M)$ and $\ln(1+q^2/M^2)+2\ln(M/\kappa)$.   The extra logarithmic term $\ln(M/\kappa)$  is obtained by treating the scale $\kappa$ as ultraviolet cutoff in the dispersion integral. The analytical continuation of the $(n-1)$-loop function to positive values of $s$ is given for the real part by
\begin{eqnarray} \text{Re}\,J_n(s) &=& {1 \over (4\pi)^{n-1} (n-2)!} \bigg\{ A_n(-s, M) \, {1\over 2\sqrt{s}} \ln{M+\sqrt{s} \over| M-\sqrt{s}|}\nonumber \\ && \qquad \qquad  \qquad \quad +L_n(-s,M) \bigg[ \ln\Big| 1- {s \over M^2}\Big|+2\ln{M\over \kappa}\bigg]+ \text{ polyn}\bigg\} \,. \end{eqnarray}
\section{Extension to 4+1 space-time dimensions}
For curiosity, let us investigate the analogous situation with $n$-body phase-spaces in $4+1$ space-time dimensions. The two-body phase-space with final-state particles of masses $m_1$ and $m_2$ is now calculated as follows:

\begin{eqnarray}
 \widetilde \Gamma_2(\mu; m_1, m_2)&=& \int{d^4k_1 \over (2\pi)^4 2\omega_1}\int{d^4k_2 \over (2\pi)^4 2\omega_2} \,(2\pi)^5 \delta^{(5)}\big(\vec P -\vec k_1-\vec k_2\big) \nonumber \\ &=& {1\over 16 \pi}\int_0^\infty dk{k^3 \over \sqrt{m_1^2+k^2} \sqrt{m_2^2+k^2}}\, \delta\Big(\mu - \sqrt{m_1^2+k^2}- \sqrt{m_2^2+k^2}\,\Big)\,,\end{eqnarray}
with $\vec P = (\mu, 0,0,0,0)$ in the center-of-mass frame. Dividing by  the derivative of the argument of the $\delta$-function with respect to $k$ leaves as an integrand $k^2/\mu$, and one inserts the solution (i.e. the center-of-mass-momentum)
\begin{equation}
k(\mu) = {1\over 2\mu} \sqrt{\lambda(\mu^2,m_1^2,m_2^2)}\,, 
\end{equation}
with $\lambda(a,b,c) = a^2+b^2+c^2-2ab-2ac -2bc$ the conventional K\"allen function. As a result the integrated two-body phase-space in $4+1$ space-time dimensions is given by the rational expression:
\begin{equation}
\widetilde \Gamma_2(\mu; m_1, m_2)= {\lambda(\mu^2,m_1^2,m_2^2) \over 64\pi \mu^3}\,, 
\end{equation}
which grows linearly ($\sim \mu-m_1-m_2$) from threshold onwards. Note that $\lambda(\mu^2,m_1^2,m_2^2)=(\mu-m_1-m_2)(\mu-m_1+m_2)(\mu+m_1-m_2)(\mu+m_1+m_2)$ depends separately on the sum $m_1+m_2$ and the difference $|m_1-m_2|$ of the two involved masses.  Using the previous result as twice the imaginary part, one finds for the one-loop function in the euclidean domain $s=-q^2$:
 \begin{eqnarray}
 \widetilde J_2(-q^2) &=& {1\over 64\pi^2} \bigg\{ (m_1+m_2)\bigg[{(m_1-m_2)^2 \over q^2}-1\bigg] \nonumber \\ && -{1\over q^3} \big[(m_1+m_2)^2+q^2\big] \big[(m_1-m_2)^2+q^2\big]\arctan{q \over m_1+m_2} \bigg\}\,.
 \end{eqnarray}
 Here, a linear divergence has been dropped, and this leads  to the value at $q^2=0$:
 \begin{equation}
\widetilde J_2(0)= -{m_1^2+m_1m_2+m_2^2 \over 24\pi^2(m_1+m_2)} \,,
\end{equation} 
that agrees perfectly with the result of dimensional regularization
\begin{eqnarray}
\widetilde J_2(0)&=&(4\pi)^{-d/2} \Gamma\Big(2-{d\over 2}\Big) \int_0^1 dx \big[m_1^2 x +m_2^2(1-x)\big]^{d/2-2}\,\big|_{d=5}\nonumber \\ &=&  -{1\over 16\pi^2}\int_0^1 dx \sqrt{m_1^2 x +m_2^2(1-x)} = -{m_1^3-m_2^3\over 24 \pi^2(m_1^2-m_2^2)}\,.
\end{eqnarray}
Next, one calculates the integrated three-body phase space in $4+1$ dimensions recursively from the two-body phase-space as
\begin{equation}
\widetilde \Gamma_3(\mu;m_1,m_2,m_3) = {1\over \pi} \int_{m_1+m_2}^{\mu -m_3}\!\! d\mu' \, \mu' \,{\lambda(\mu^2,\mu'^2,m_3^2) \over 64\pi \mu^3}\,{\lambda(\mu'^2,m_1^2,m_2^2) \over 64\pi \mu'^3} \,,
\end{equation}
and obtains the result
\begin{equation}
\widetilde \Gamma_3(\mu;m_1,m_2,m_3) = {(\mu -\sigma_1)^3\over 105 (8\pi \mu)^3} \Big\{\mu^4+3\mu^3 \sigma_1 +4\mu^2(7\sigma_2 -2 \sigma_1^2) +(3\mu+\sigma_1)(\sigma_1^3-7\sigma_1\sigma_2 +35 \sigma_3) \Big\} \,,
\end{equation}
with 
\begin{equation} \sigma_1 = m_1+m_2+m_3\,, \qquad \sigma_2=m_1m_2+m_1m_3+m_2m_3\,, \qquad \sigma_3 =m_1m_2m_3\,,
\end{equation}
the elementary symmetric polynomials in the three masses $m_1, m_2, m_3$. The same result as in eq.(31) can be obtained by following eq.(3), performing angular integrals, and arriving at an integration over the Dalitz region in the $\omega_1\omega_2$-plane: 
\begin{equation} \widetilde \Gamma_3(\mu;m_1,m_2,m_3) ={1\over 128\pi^4}\iint_{D(\omega_1,\omega_2)> 0}\!\!\! d\omega_1 d\omega_2 \sqrt{D(\omega_1,\omega_2)}\,. \end{equation} 
In the next step the intermediate result  $\int_{\omega_2^-}^{\omega_2^+}\! d\omega_2\sqrt{(\omega_2^+ -\omega_2) (\omega_2-\omega_2^-) } = \pi(\omega_2^+ -\omega_2^-)^2/8$ is used and with this the integral $\int_{m_1}^{\omega_1^\text{end}} \! d\omega_1 \sqrt{\mu^2-2\mu \omega_1+m_1^2} (\omega_2^+ -\omega_2^-)^2$ can be  performed to reproduce the expression in eq.(31).
 
The partial fraction decomposition of $\mu \widetilde \Gamma_3(\mu;m_1,m_2,m_3)/(\mu^2+q^2)$ into pieces proportional to $1/(\mu^2+q^2)$ and $-2\mu/(\mu^2+q^2)$ plus a polynomial allows to infer the two-loop function $\widetilde J_3(-q^2)$ as a linear combination of $q^{-1}\arctan(q/\sigma_1)$ and $\ln(1+q^2/\sigma_1^2)+2\ln(\sigma_1/\kappa)$.
Modulo additive polynomial terms $(\sim 1, q^2, q^4)$, this two-loop function reads  
\begin{eqnarray}
\widetilde J_3(-q^2)&\!\!\!=\!\!\!& {1\over 6(4\pi)^4} \bigg\{\bigg[ {\sigma_1^4\over q^2}\Big({\sigma_1^3 \over 35}\!-\!{\sigma_1 \sigma_2\over 5} \!+\!\sigma_3\Big)\!+\! 2\sigma_1^2 \Big({\sigma_1^3 \over 5}\! -\!\sigma_1 \sigma_2 \!+\!3\sigma_3\Big)+(3\sigma_1 \sigma_2\!-\!\sigma_1^3\!-\!3\sigma_3)q^2\bigg]{1\over q}\arctan{q \over \sigma_1} \nonumber \\ && + \bigg[ \sigma_1\Big({\sigma_1^3 \over 2} -2\sigma_1 \sigma_2 +4\sigma_3\Big) +{q^2 \over 5}(2\sigma_2-\sigma_1^2) -{q^4\over 70}\bigg]\bigg[\ln\Big(1+ {q^2\over \sigma_1^2}\Big) + 2\ln{\sigma_1 \over \kappa}\bigg]\nonumber \\ && + {\sigma_1^3 \over q^2}\Big( {\sigma_1 \sigma_2\over 5} -\sigma_3 -{\sigma_1^3 \over 35}\Big)+\text{polyn}\bigg\} \,.
\end{eqnarray}
Note the extra $1/q^2$-term in third line, which  makes the whole expression singularity-free at $q^2=0$.

The next iterative step to the four-body phase space is straightforward by employing eq.(6) and this leads to the following result for the integrated four-body phase-space in $4+1$ dimensions:
\begin{eqnarray}&& \widetilde \Gamma_4(\mu;m_1,m_2,m_3,m_4) = {(\mu -\Sigma_1)^5\over 630 (8\pi )^5 \mu^3} \bigg\{{\mu^5 \over 10} +{\mu^4\over 2}\Sigma_1 +\mu^3(7\Sigma_2 -2 \Sigma_1^2)\nonumber \\ && \qquad +
\mu^2 (2\Sigma_1^3-13\Sigma_1\Sigma_2 +48 \Sigma_3)+(5\mu+\Sigma_1)\bigg[\Sigma_1^2 \Sigma_2-{\Sigma_1^4 \over 10} -8\Sigma_1\Sigma_3+56\Sigma_4\bigg] \bigg\} \,,
\end{eqnarray}
with the abbreviations
\begin{eqnarray}
&& \Sigma_1 = m_1+m_2+m_3+m_4\,,\nonumber \\ && 
 \Sigma_2=m_1m_2+m_1m_3+m_1m_4+m_2m_3+m_2m_4+m_3m_4\,,\nonumber \\ &&   \Sigma_3 =m_1m_2m_3+m_1m_2m_4+ m_1m_3m_4+m_2m_3m_4\,,\nonumber \\ && \Sigma_4 = m_1m_2m_3m_4 \,,\end{eqnarray}
for the elementary symmetric polynomials in the four masses $m_1, m_2, m_3,m_4$. The associated three-loop function $\widetilde J_4(-q^2)$ has a form analogous to eq.(34) with the extension that the square bracket in front of the basic function  $q^{-1}\arctan(q/\Sigma_1)$ includes now additional polynomial terms proportional to $q^4, q^6, q^8$. 

In the next iterative step, one finds that the  five-body phase-space in $4+1$ dimensions takes the following form 
\begin{eqnarray} \widetilde \Gamma_5(\mu;m_1,\dots,m_5) & =& {2(\mu - \tilde \Sigma_1)^7\over 22275 (8\pi)^7 \mu^3} \bigg\{ {\mu^6 \over 91} +{\mu^5 \over 13} \tilde \Sigma_1 +{\mu^4 \over 7}\bigg( 10 \tilde \Sigma_2 -{37\over 13}\tilde \Sigma_1^2\bigg) \nonumber \\ & & + {\mu^3\over 7}\bigg( {58\over 13} \tilde \Sigma_1^3-29\tilde \Sigma_1\tilde \Sigma_2 +99 \tilde \Sigma_3\bigg) +{\mu^2\over 7}\bigg( 880 \tilde \Sigma_4-187 \tilde \Sigma_1\tilde \Sigma_3 +27 \tilde \Sigma_1^2 \tilde \Sigma_2-{37 \over 13} \tilde \Sigma_1^4\bigg) \nonumber \\ && +\bigg( \mu +{\tilde \Sigma_1\over 7}\bigg)\bigg( {\tilde \Sigma_1^5 \over 13} - \tilde \Sigma_1^3\tilde \Sigma_2 +11\tilde \Sigma_1^2 \tilde \Sigma_3-110 \tilde \Sigma_1\tilde \Sigma_4+990\tilde \Sigma_5\bigg) \bigg\} \,, 
\end{eqnarray} 
with the abbreviations
\begin{eqnarray}
&& \tilde \Sigma_1 = m_1+m_2+m_3+m_4+m_5\,,\nonumber \\ && 
 \tilde \Sigma_2=m_1m_2+m_1m_3+m_1m_4+m_1m_5+m_2m_3+m_2m_4+m_2m_5+m_3m_4+m_3m_5+m_4m_5\,,\nonumber \\ &&   \tilde \Sigma_3 =m_1m_2m_3+m_1m_2m_4+ m_1m_2m_5+m_1m_3m_4+ m_1m_3m_5\nonumber \\ && \qquad\,\,  +m_1m_4 m_5+m_2m_3m_4+ m_2m_3m_5+m_2m_4m_5 + m_3m_4m_5 \,,\nonumber \\ && \tilde \Sigma_4 = m_1m_2m_3m_4+  m_1m_2m_3m_5 +m_1m_2m_4m_5+ m_1m_3m_4m_5+ m_2m_3m_4m_5\,,\nonumber \\ && \tilde \Sigma_5 = m_1m_2m_3m_4m_5\,, 
 \end{eqnarray}
for the elementary symmetric polynomials in the five masses $m_1, m_2, m_3,m_4, m_5$. By induction, one can deduce that the $n$-body phase-space takes the form
$\widetilde \Gamma_n(\mu;m_1,\dots,m_n)= (8\pi)^{3-2n} \mu^{-3}\big(\mu-\sum_{j=1}^nm_j\big)^{2n-3} P_{n+1}(\mu)$, where the last factor $P_{n+1}(\mu)$ is a polynomial in $\mu$ of degree $n+1$, whose coefficients are totally symmetric in the $n$ masses $m_1,\dots,m_n$.

\section{Digression to 1+1 dimensions}
In passing we note that the two-body phase-space in $1+1$ space-time dimensions is given by the expression
\begin{equation}
\bar \Gamma_2(\mu;m_1,m_2) = {1 \over \sqrt{\lambda(\mu^2,m_1^2,m_2^2)}} \,,
\end{equation} 
which exhibits a singular behavior $\sim (\mu-m_1-m_2)^{-1/2}$ near threshold. The associated one-loop function in $1+1$ dimensions is then calculated as
\begin{equation}\bar J_2(-q^2) = {1\over \pi W_+ W_-} \ln{ W_+ +W_- \over 2\sqrt{m_1 m_2}}\,, \quad \qquad W_\pm = \sqrt{q^2+(m_1\pm m_2)^2}\,. \end{equation}
The iterative procedure (see eq.(6)) leads for the three-body phase-space in $1+1$ dimensions to the integral representation
\begin{equation}
\bar \Gamma_3(\mu;m_1,m_2,m_3) = {1 \over 2\pi} \int_{(m_1+m_2)^2}^{(\mu-m_3)^2} ds' \,\big[\lambda(\mu^2,s',m_3^2)\,\lambda(s',m_1^2,m_2^2)\big]^{-1/2} \,,
\end{equation}
which gives rise to a non-vanishing value at threshold
\begin{equation}
\bar \Gamma_3(m_1+m_2+m_3;m_1,m_2,m_3) = {1 \over  8 \sqrt{(m_1+m_2+m_3)m_1m_2m_3}}= {1\over 8 \sqrt{\sigma_1\sigma_3}}\,.
\end{equation}
Actually, the integral representation of $\bar \Gamma_3(\mu;m_1,m_2,m_3)$ can be mapped by a linear fractional transformation onto that of a complete elliptic integral of the first kind:
\begin{eqnarray}
\bar \Gamma_3(\mu;m_1,m_2,m_3) &=& {1 \over 8\pi\sqrt{\mu m_1m_2m_3}} \int_1^\rho\! dx {1\over \sqrt{x(x-1)(\rho-x)} }\nonumber \\ &=&{1 \over 4\pi\sqrt{\mu \sigma_3}} \int_0^{\pi/2}\! d\varphi \big[1 +(\rho-1)\sin^2\varphi\big]^{-1/2}\,,
\end{eqnarray}
with the auxiliary parameter
\begin{eqnarray}
\rho &=& {1\over 16 \mu m_1m_2m_3}\big[(\mu+m_3)^2-(m_1+m_2)^2\big] \big[(\mu-m_3)^2-(m_1-m_2)^2\big]\nonumber \\ &=& {1\over 16 \mu \sigma_3 }\big[ \mu^4+2\mu^2(2\sigma_2-\sigma_1^2)+8 \mu \sigma_3 
+\sigma_1^4 -4 \sigma_1^2\sigma_2+8 \sigma_1 \sigma_3\big] >1 \,,
\end{eqnarray}
that is totally symmetric in the three masses $m_1,m_2,m_3$. 
The $\rho$-dependent value of the complete elliptic integral $\int_1^\rho dx[x(x-1)(\rho-x)]^{-1/2}$ in eq.(43) turns out to be equal to the real Weierstra{\ss} half-period $\Omega_1/2$ that belongs to the two (real) Weierstra{\ss} invariants
 \begin{equation}
 g_2 =\sum_{n_1,n_2}\!\!\,'{60 \over (n_1 \Omega_1 +i n_2 \Omega_2 )^4} = {\rho^2-\rho+1\over 12}\,, \qquad g_3 = \sum_{n_1,n_2}\!\!\,'{140 \over (n_1 \Omega_1 +i n_2 \Omega_2 )^6}= { (2 - \rho) (1+\rho) (2\rho - 1)\over 432}\,.
\end{equation}
Note that the other Weierstra{\ss} half-period $i\, \Omega_2/2$ is purely imaginary, and thus the period lattice of the underlying doubly-periodic elliptic function in the complex plane is of rectangular shape.  The transformation of the cubic polynomial under the square root in eq.(43) to the Weierstra{\ss} normal form $4 \wp(z)^3-g_2 \wp(z)-g_3$ is achieved by setting $x = -4 \wp(z)+(1+\rho)/3$.

\end{document}